# Effect of Sr substitution on the stability of CuO$_{1-\delta}$ chains in oxygen and nitrogen annealed YBa$_{2-x}$Sr$_x$Cu$_3$O$_{7-\delta}$ (x = 0.0, 0.50, 1.0) superconducting system

M.A. Ansari[*], V.P.S. Awana, Anurag Gupta, R.B. Saxena, and H. Kishan

National Physical Laboratory, K.S. Krishnan Marg, New Delhi 110012, India.

Samples of YBa$_{2-x}$Sr$_x$Cu$_3$O$_{7-\delta}$ (x = 0.0, 0.50, 1.0) series were synthesized by solid-state reaction and first annealed in flowing O$_2$ gas. All samples crystallized in orthorhombic structure, where orthorhombic distortion was found to decrease with an increase in x. Four probe resistivity measurements showed that superconducting transition temperature (T$_c$) decreases monotonically with x at a rate of dT$_c$/dx = 1.3 K/at%. The sample with x=0 had a T$_c$ = 91 K. The samples were further annealed in flowing N$_2$ gas at 450 $^0$C. The T$_c$ of all samples decreased by nitrogen annealing. Interestingly, the decrease in T$_c$(x) became nearly three times larger at a rate of dT$_c$/dx = 3.7 K/at%. The sample with x=0 had a T$_c$ = 42 K. We discuss possible reasons and implications of the T$_c$(x) suppression vis-à-vis stability of CuO$_{1-\delta}$ chains.

*Presenting /corresponding Author: E-mail ansari@mail.nplindia.ernet.in

## 1. INTRODUCTION

The RE-123 structure can be viewed as (Ba)O / CuO$_2$ / RE / CuO$_2$ / (Ba)O slabs interconnected by a Cu-O sheets containing chains of Cu and O atoms with variable composition CuO$_{1-\delta}$ [1]. The oxygen sites in the CuO$_2$ planes are identified as O(2) and O(3). The oxygen site in the SrO plane is called O(4). The RE plane is devoid of any oxygen. In general oxygen sites O(2), O(3), and O(4) are always fully occupied and any intake or release of oxygen takes place only from CuO$_{1-\delta}$ chains. The amount of oxygen released and as a result the decrease in T$_c$ (superconducting transition temperature) depends not only on external reducing parameters but the intrinsic stability of CuO$_{1-\delta}$ chains.

The substitution of Sr at Ba-site though decreases marginally the superconducting transition temperature ($T_c$) of the RE-123 [2], it increases substantially the performance of the compound under magnetic field, or otherwise improves the irreversibility line characteristics [3]. Under normal synthesis conditions Ba can be substituted only upto 50 % by Sr, for full substitution one needs to employ the high pressure high temperature route [4]. In the present study we study the change in the structure and $T_c$ of the $YBa_{2-x}Sr_xCu_3O_{7-\delta}$ (x = 0.0, 0.50, 1.0) samples with respect to the removal of oxygen from chains. Our results show that the release of oxygen from $YBa_{2-x}Sr_xCu_3O_{7-\delta}$ compound increases with x, confirming the fact that Sr substitution at Ba-site introduces instability in Cu-O chains of RE-123 structure.

## 2. EXPERIMENTAL DETAILS

Samples of $YBa_{2-x}Sr_xCu_3O_{7-\delta}$ (x = 0.0, 0.50, 1.0) series were synthesized by solid-state reaction using ingredients $Y_2O_3$, $SrCO_3$, $BaCO_3$ and CuO. Calcinations were carried out on mixed powders at 880 $^0$C, 890 $^0$C, 900 $^0$C and 910 $^0$C each for 24 hours with intermediate grindings. Two set of pressed pellets were annealed in a flow of oxygen at 920 $^0$C for 40 hours and subsequently cooled slowly to room temperature with an intervening annealing for 24 hrs. at 600 $^0$C. On one set of pellets called "$O_2$-annealed" X-ray diffraction (XRD) data were obtained at room temperature (MAC Science: MXP18VAHF[22]; Cu$K_\alpha$ radiation) and four probe resistance (R) measurements in the temperature (T) range of 12 - 300 K, were performed in a close cycle refrigerator. The other set of pellets was further annealed in flowing $N_2$ gas at 450 $^0$C for 24 hrs., and will be called as "$N_2$-annealed". R(T) measurements were also performed on these samples.

## 3. RESULTS AND DISCUSSION

X-ray diffraction (XRD) patterns of all the $O_2$-annealed $YBa_{2-x}Sr_xCu_3O_{7-\delta}$ (x = 0.0, 0.50, 1.0 samples are shown in Fig.1. All the samples studied are phase pure in nature, except for x = 1.0 sample where small unidentified impurity lies in the compound. Worth mentioning is the fact that through normal synthesis routes (without applying HPHT process) x = 1.0 is the solubility limit of Sr at Ba site in $YBa_{2-x}Sr_xCu_3O_{7-\delta}$ compounds. Lattice parameters determined from XRD data revealed that these samples crystallize in an orthorhombic structure. The lattice parameters are: $a$ = 3.824(2) Å, $b$ = 3.892(1) Å, and $c$ = 11.672(3) Å, for x = 0.0; $a$ = 3.806(4) Å, $b$ = 3.869(3) Å, and $c$ = 11.606(3) Å, for x = 0.50 and $a$ = 3.805(2) Å, $b$ = 3.856(3) Å, and $c$ = 11.572(5) Å, for x = 1.0. Substantial decrease in c-axis lattice parameter indicates successful Sr substitution at Ba-site in $YBa_{2-x}Sr_xCu_3O_{7-\delta}$. Orthorhombic distortion, i.e. *b-a,* decreases with x, indicating a decrease of oxygen content from Cu-O chains in the structure with Sr substitution, which is in agreement with previous reports [2,3].

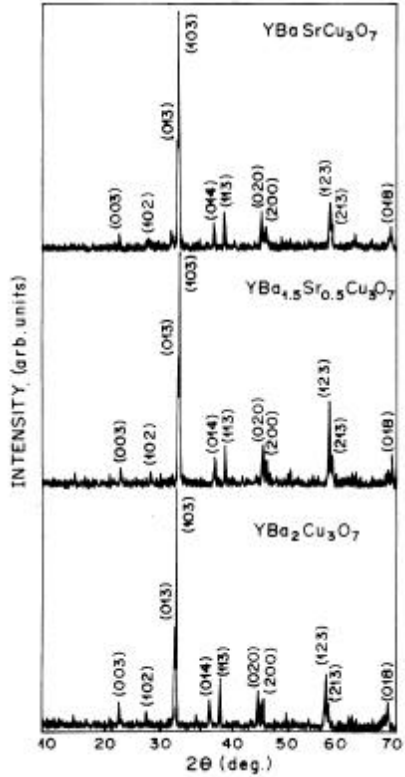

Fig.1 X-ray diffraction patterns for $O_2$-annealed $YBa_{2-x}Sr_xCu_3O_{7-\delta}$ samples

Fig. 2 shows the resistance versus temperature plots for both $O_2$- and $N_2$- annealed $YBa_{2-x}Sr_xCu_3O_{7-\delta}$ compounds in temperature range of 12 to 200 K. In $O_2$-annealed $YBa_{2-x}Sr_xCu_3O_{7-\delta}$ samples the superconducting transition temperature decreases monotonically with x. Decrease of $T_c$ with x in $YBa_{2-x}Sr_xCu_3O_{7-\delta}$ is in agreement with previous report [2,3]. Quantitatively the decrease in $T_c$ ($dT_c/dx$) is 1.3K/at% of Sr substituted at Ba-site in $YBa_{2-x}Sr_xCu_3O_{7-\delta}$.

It was expected that substitution of bigger Ba cation by smaller cation Sr in RE-123 structure could increase the $T_c$ by providing the internal crystallographic pressure. However this does not happen and ironically the $T_c$ decreases with x in $YBa_{2-x}Sr_xCu_3O_{7-\delta}$ system. However, note that the internal crystallographic pressure probably does get applied with Sr substitution, being evidenced by decrease in c-lattice parameter, but the same is probably not uniform over the unit cell. In particular the distance between the two superconducting $CuO_2$ planes increases instead of decreasing with x in $YBa_{2-x}Sr_xCu_3O_{7-\delta}$ [4,5].

This explains why $T_c$ decreases with x in $YBa_{2-x}Sr_xCu_3O_{7-\delta}$ system. Another fact evidenced by decrease in orthorhombic distortion by increase in x is that oxygen content decreases with x in $YBa_{2-x}Sr_xCu_3O_{7-\delta}$ system. We believe that the latter fact of decreased oxygen content coupled with the decreasing internal pressure on $Cu-O_2$ superconducting planes are both responsible for decrease of $T_c$ in $YBa_{2-x}Sr_xCu_3O_{7-\delta}$ system.

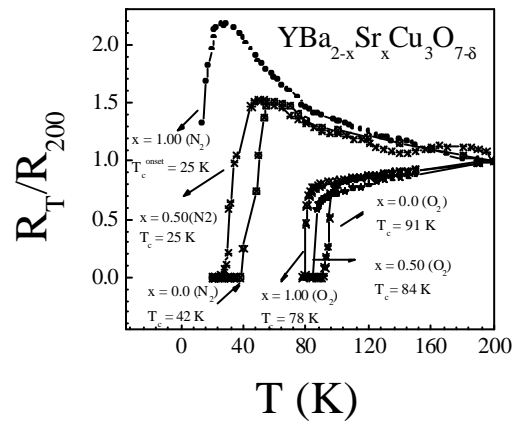

Fig. 2 Resistance versus temperature plots for both $O_2$ and $N_2$ annealed $YBa_{2-x}Sr_xCu_3O_{7-\delta}$ samples

Decrease in oxygen content of $YBa_{2-x}Sr_xCu_3O_{7-\delta}$ with x reveals that Sr substitution increases the instability of $CuO_{1-\delta}$ chains of the normal RE-123 structure and makes them more susceptible to the release of oxygen. The results of resistance versus temperature for $N_2$-annealed $YBa_{2-x}Sr_xCu_3O_{7-\delta}$ compounds are shown in Fig.2. The $T_c$ for x = 0 (pristine) sample decreases from 91 K to 42 K. For x = 0.50 and 1.0 samples the $T_c$ decreases from 84 K and 78 K to 25 K and around 5 K (extrapolated) respectively after $N_2$-annealing at 450 $^0$C. Quantitatively the decrease in $T_c$ ($dT_c/dx$) is 3.7 K/at% of Ba site Sr substitution in $N_2$-annealed $YBa_{2-x}Sr_xCu_3O_{7-\delta}$. Interestingly the $dT_c/dx$ for $N_2$-annealed samples is 3 times (3.7 K/at%) when compared to oxygen annealed (1.3 K/at%) $YBa_{2-x}Sr_xCu_3O_{7-\delta}$. This shows that Sr substitution at Ba-site decreases $T_c$ much faster in oxygen deficient $YBa_{2-x}Sr_xCu_3O_{7-\delta}$ system than in oxygenated compounds. Before practical applications of $YBa_{2-x}Sr_xCu_3O_{7-\delta}$ compounds, due to their better irreversibility line characteristics [3], one should consider the relative instability of $CuO_{1-\delta}$ chains in them at various Sr content. Conclusively, our results show that stability of the $YBa_{2-x}Sr_xCu_3O_{7-\delta}$ compounds towards release of oxygen decreases with x.